# How constraints on editing affects cultural evolution

Ofer Tchernichovski[1], Eitan Globerson[2,3], Peter Harrison[4] & Nori Jacoby[5]

1. otcherni@hunter.cuny.edu, Department of Psychology, Hunter College, CUNY New York, NY 10065 USA. 2. eitangl@technion.ac.il Technion: Israel Institute of Technology. Technion City, Haifa 3200003, Israel. 3. Jerusalem Academy of Music and Dance, Givaat Ram Campus, Jerusalem. 4. pmch2@cam.ac.uk Centre for Music and Science, University of Cambridge, Cambridge UK. 5. kj338@cornell.edu, Department of Psychology, Cornell University Ithaca, NY 14853 USA

**Abstract**

When is it beneficial to constrain creativity? Creativity thrives with freedom, but when people collaborate to create artifacts, there is tension between giving individuals freedom to revise, and protecting prior achievements. To test how imposing constraints may affect collective creativity, we performed cultural evolution experiments where participants collaborated to create melodies and images in chains. With melodies, we found that limiting step size (number of musical notes that can be changed) improved pleasantness ratings for created tunes. Similar results were observed in cohorts of musicians, and with different selection regimes. In contrast, limiting step size in creating images consistently reduced pleasantness. These conflicting findings suggest that in domains such as music, where artifacts can be easily damaged, and where evolutionary outcomes are hard to foresee, collective creativity may benefit from imposing small step sizes. We discuss parallels with search algorithms and the evolution of conservative birdsong cultures.

## Introduction

Collective creativity is fundamental to human culture (Giuffre, 2016). While creativity can sometimes be the product of an individual who single handedly and rapidly transforms culture—figures like Isaac Newton, Leonardo da Vinci, Ludwig van Beethoven, and Aristotle exemplify this—cultural change is often more gradual, emerging from the accumulated contributions of many individuals over time (Mesoudi & Thornton, 2018).

The study of cumulative cultural evolution examines how small, incremental contributions can lead to technological advancements and problem-solving innovations(Derex et al., 2019; Mesoudi & Thornton, 2018). In the arts, vernacular musical styles have historically evolved through the collective contributions of individuals (Savage et al., 2022), often through the independent development and recombination of stylistic features to form new musical traditions (Yoel, 2022).

This highlights an inherent tension between abrupt creative shifts and gradual cultural evolution. While sudden breakthroughs can propel culture forward, excessive disruption may erase prior achievements and hinder long-term progress. Conversely, gradual change preserves past innovations but can be overly conservative. When and why does collective creativity benefit from constraints?

Traditional assessments of creativity, such as those by Torrance (1966), Guilford (1967), and Boden (1990), often test individuals in settings designed to maximize creative output, where higher creativity scores are associated with generating as many ideas as possible. However, in practice, constraints can paradoxically enhance creativity. For instance, practitioners frequently report that brainstorming sessions become more productive when constraints are introduced (Gould et al., 2023; Ulibarri et al., 2019).

The dynamics of individual creativity reveal a balance between exploration and exploitation. In tasks like designing aesthetically appealing visual displays, individuals alternate between small, incremental refinements (exploitation) and larger, more radical changes (exploration) (Hart et al., 2018). But how does this trade-off apply to collective creativity? Under what conditions do constraints enhance or hinder creative collaboration?

To investigate this, we take an empirical approach, building on the tradition of using transmission chains to study cultural evolution (Kirby et al., 2014; Niarchou et al., 2022). Research on both humans and other animals has shown that transmission chains shape evolving patterns through iterative imitation (Claidière et al., 2014; Claidière & Sperber, 2024; Fehér et al., 2009; Singleton & Newport, 2004).

Here, we examine the creative process using interactive interfaces for artistic design (music and visual art). We introduce varying levels of constraints in the transmission process to observe their effects on the cultural artifacts that were produced. The first study empirically tests additional mechanisms of cultural evolution, potentially including interactions between selection and transmission.

In order to test these questions experimentally, we deployed asynchronous online experiments using Unity 3D and PsyNet (Harrison et al., 2020; Tchernichovski et al., 2023), where participants collectively created artifacts in chains. We tested how imposing different types of constraints affect the pleasantness scores of the evolutionary outcomes, which we assessed independently. We tested for an effect of two types of constraints. One type is constraints imposed on the creative process, affecting the similarity between successive iterations. The second type of constraints are imposed on the transmission process, setting selection mechanisms through which artifacts propagate to the next iteration (generation) of the chain (Boyd & Richerson, 1988).

## Methods

*Participants*

We recruited a total of 1466 participants that provided consent in accordance with an approved protocol. All participants were recruited online using Prolific and Amazon Mechanical Turk (MTurk) with the following constraints on recruitment: (i) participants must be at least 18 years old, and (ii) have a 95% or higher approval rate on previous tasks. Participants were compensated at a rate of USD $11 per hour, with payment based on the portion of the experiment

completed. The full experiment took approximately 10–25 minutes to complete.

*Recruitment of musicians*

Musicians were recruited through flyers, emails, and word of mouth among professional musicians and music students. Participants included a mix of local musicians from a music academy in Frankfurt and musicians from across Europe. On average, they had X years of daily practice on their instrument. Musicians were enrolled in a designated recruitment system (Lee et al. 2021) and, after providing consent, could select from various available tasks, including the current study.

*Games design*

All games were coded using C# scripts in Unity 3D, combined with python code in PsyNet (https://www.psynet.dev/), a Python package for performing complex online behavioral experiments at large scale ( Harrison et al., 2020). The Psynet code controlled the game logic through Unity (compiled to WebGL), and managed the online implementation at the client side via Chrome web browser. Devices for creating artifacts (synthesizers, draw canvases) were coded in C# Unity as well. At the client side, the app ran via Chrome web browser, which communicated with a backend Python server cluster responsible for organizing the experiment and collecting data.All games were deployed asynchronously, such that changes in artifacts posted by previous players accumulated in the world that the next players encountered.

**Step synthesizer design**

We run experiments where participants created and edited simple melodies (tunes) using a step synthesizer (**Fig. 1A**).. The synthesizer has 16 time slots and three note types (grand piano middle C,D,E) that can be triggered individually or simultaneously or in chords. The three musical notes were synthesized from a commercial piano synthesizer. Time slots can be used to insert gaps as well. A 'play' button allowed creators to listen to their tune at all stages. Although the synthesizer can only generate short and simple tunes of 3.2 seconds, the options add up to a massive space of $8^{16}$ (~281 trillion) possible unique melodies.

**Draw canvas design**

We coded a canvas in Unity 3D. We created two types of canvas: A monochromatic canvas consists of a 30x30 pixel board. Each pixel started as white, and is toggled to black (and back to white) by pointing and clicking the mouse. The second type of drawing canvas was tri-chromatic. Its size was much smaller (8x8 pixels). In each implementation, we grayed out a random set of 16 pixels, and denoted them as unchangeable (this makes the creation of images more challenging). All other pixels were initially white, turned black on the first click, yellow on a second click and reset back to white on a third click.

*Transmission chains network assignment*

The experiment involved constructing recursive 'chains' of evolving stimuli. In each trial, a 'creator' participant was randomly assigned to a chain. The creator was given a source artifact (or a blank) and used an interface (a synthesizer or a canvas) to create a derived artifact. Creators were given the task of making the artifact more appealing, by either adding, removing or changing elements (musical notes or pixels). We required participants to spend at least one minute creating each artifact. Participants were allowed to explore different options (using the 'reset' button) before submitting their final version of the artifact. Creators could not contribute more than a single artifact per chain. Once done, PsyNet moved the creator to a different chain. In this manner, each participant created several (usually 5-8) artifacts.

**Artifact creation rules**

All artifacts were created recursively in a chain. Within a chain, in each step, the artifact was created/edited by a unique participant.

Step limit rule (**Fig. 1B**) sets an upper limit to the overall number of changes a creator can make in each turn by adding, shifting or removing elements. We did not limit exploration: regardless of step limit, creators were allowed to start over (reset) as often as needed.

Overcrowding limit rule (**Fig. 1C**) limits the overall number of elements in the artifact. It was only implemented only for melodies. When turned on, the maximal number of musical notes in the tune was set to five A, five B and five C notes (15 notes overall), preliminary pilot studies indicated a tendency for adding too many notes, making the tunes unpleasant. which we empirically determined to be optimal. When turned off, there was no set limit on the overall number of notes in the tune.

**Artifact propagation rules (Create & Rate paradigm)**

To simulate a cultural system incorporating both selection and imitation, we employed a Create & Rate paradigm, where artifacts competed against one another in a selection process that determined which would propagate to the next node in the chain. Participants were randomly assigned to the role of a creator or a rater. Each chain begins by assigning the first node with a 'create' state, and a creator is recruited to generate an artifact. After the creator submitted the artifact, the chain transitioned to a 'rate' state, and raters were recruited. Raters were then presented with two candidate artifacts and selected the one they found more appealing. After each candidate was rated seven times (score = 0-7), the candidate with higher score was propagated to the next node, and the chain transitioned to a 'create' state, and so forth. With the Create & Rate paradigm, we experimented with two types of melody selection.

Local selection rule (**Fig. 1D**): In each step, newly created tunes were evaluated against their source within a chain. That is, the two candidates (source tune and derived tune) were judged against each other. This selection rule induces priming in the ratings.

Global selection rule (**Fig. 1D**). In each step, each candidate tune (source tune and derived tune) was rated against a randomly sampled tune from other chains. In this manner, the tunes were rated against the background of the entire space of created tunes, which prevents priming.

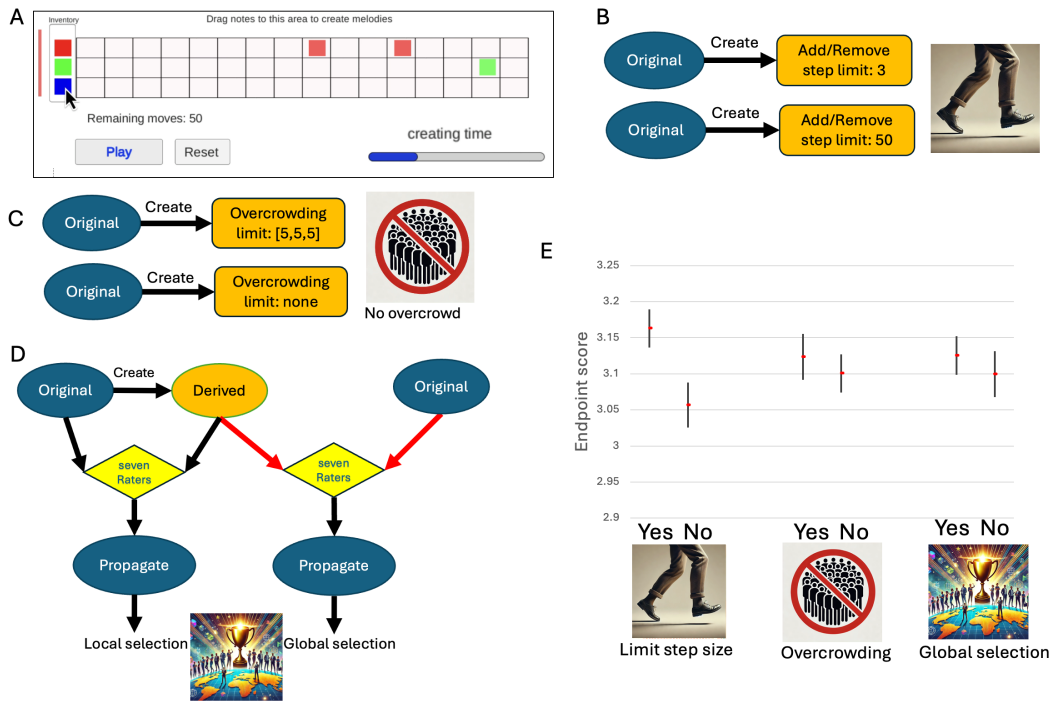

**Figure 1**: A, a step-synthesizer GUI for creating and editing simple Do-Re-Mi melodies. B, constraints on step size (number of musical notes that can be changed). C, Constraints on overall number of notes in a melody (crowdedness). D, Darwinian selection constraints (local vs. global competition). E. Mean pleasantness scores obtained at the end of the chain (after 20 iterations). Error bars are s.e.m for the marginal distributions of ratings. Sample size for Create & Rate: 853 participants; 32 chains; 5120 melodies created; 30,720 ratings. Sample size for validation (pleasantness score): 10 independent ratings per melody.

**Validation experiments**

To assess outcomes, after each experiment we recruited an additional cohort of participants to rate the created artifacts. Participants were presented with artifacts in random order (40 artifacts per participant), who rated the pleasantness of each artifact separately, on a scale of 1-4. In this manner, we obtained 10 independent evaluations for each artifact, and used the mean as the pleasantness estimate.

*Statistical analysis*

For the Evolution of artifacts under different constraints experiment we fit a linear regression model:

score ~ #notes * chain_iteration * step_limit(binary) * crowding_limit (binary) * selection (binary)

For other experiments we used a simpler model:

score ~ treatment_group * iteration

## Results

**Evolution of melodies under different constraints**

We recruited 840 participants to test how imposing different types of constraints may affect the evolution of melodies under different selection regimes. For each chain, tunes evolved recursively over 20 iterations, following a Create & Rate paradigm. Our eight treatment groups included all permutations of three binary rules, with each rule imposing a different type of constraint: Step limit limited the number of changes creators could make. It was set to three steps (up to three changes) when turned on, and to 50 steps when turned off (essentially unlimited with our 48 slots synthesizer). Preliminary pilot indicated a tendency for adding too many notes, making the tunes unpleasant. To separate the influence of overcrowding from that of the step-limit, we imposed an Overcrowding limit rule. When turned off, there was no set limit on the overall number of notes in the tune. Finally, we wanted to test if constraints on the creative process may interact with information flow and selection regimes in our network. We therefore added a Selection rule **(Fig. 1D)**: When turned off, selection by raters was local (within a chain). When turned on, selection was global (across chains).

We deployed this experiment over all permutations of binary rules over 32 chains (four chains per group), and 20 iterations per chain, and obtained 5,120 melodies. **Figure 1E** presents the pleasantness scores at the endpoint, pooling results obtained for each rule (marginal distributions) across chains. As shown, the strongest effect was of step size: when turned on (three step limit), endpoint melodies were rated higher. No such an effect was observed for the endpoint marginal of the other rules. Fitting a linear regression model on our entire dataset revealed a significant effect of treatment on melodies scores (Multiple R$^2$= 0.179 ,F= 5.458 on 31 and 775 DF, p < 2.2e-16). The strongest variable affecting melody pleasantness was step size (t=-3.9, df=775, p=0.0001). As expected, the number of notes in a melody had a significant effect as well (t=-3.2, p=0.001). Selection type and overcrowded limit showed weaker effects (Overcrowding t = -2.3, p=0.01 ; Global selection t = -2.2, p=0.02). The only significant interactions were between #notes, chain iteration and step size, which is expected. Importantly, there was no interaction between the rules.

**Step-size effect in non musicians & musicians**
We tested if the step-size effect might be limited to cases where melodies are created by non-musicians. We also wanted to test if we can replicate the effect of step size when isolated from the other constraints and without any selection regime, in simple feed forward chains. We recruited a cohort of professional musicians (n=42 participants) as compared to a new general non-musician cohort (n=384 participants) recruited via MTurk. For the musician cohort, due to the small sample size of musicians, we compared two treatment groups, limiting step limits to either 3 or 50. For the non-musicians group, we tested three groups: step size 3, 6, & 50. For validation study, we recruited two cohorts (of non musicians), one for rating the melodies created by musicians and a separate cohort for rating melodies created by non-musician, to avoid bias. To sample the space of melodies more evenly then in the previous study, we seeded each chain with a random melody of different complexity, ranging between 4 and 12 notes). Participants created and edited melodies using the same step synthesizer. At the end of the experiment we obtained 12 iterations in each chain.

**Figure 2A** presents examples of melodies at the top and bottom scores. As expected, overcrowded melodies were typically scored low. In both non musicians (**Fig. 2B**) and musicians (**Fig. 2C**) results replicated those of the previous study. For non musicians, step size 3 gave higher scores compared to both 6 & 50-steps (t=-4.421, p=1.33e-05 & t=-3.02 p=0.002). As shown, even at step size 6, scores declined to similar levels to those of the step size 50 group. With musicians the effect of step size was similar, and even stronger (**Fig. 2C**, t=-5.817, p= 8.82e-09).

Looking at the dynamics of the changes in pleasantness score among the musician group, we found that in the first iteration, the 50 step-limit group performed very well (**Fig. 3D**). The advantage of the 3-steps group stemmed from a decline in the 50-step limit group scores during the later iterations. In the 3-step limit group -- which we ran in parallel with the same musicians -- the score increased slowly, and showed no decline in the later iterations. As a result, the 3-step treatment eventually outperformed the 50-step treatment. To further explore these unanticipated dynamics, we estimated how adding and removing musical notes affect the score in each iteration in the 50 step group (**Fig. 2E-F**). As shown, in iteration where musicians added a small number of notes we see more improvement in melody score (**Fig. 2E**, t=-2.9, p=0.0036). Counting the number of notes removed in each iteration reveals the opposite effect: musicians who removed more notes were more successful in improving the melody score (**Fig. 2F**, t=3.1, p=0.0022). We have seen a similar deterioration of scores in the big step size groups of other experiments (not shown).

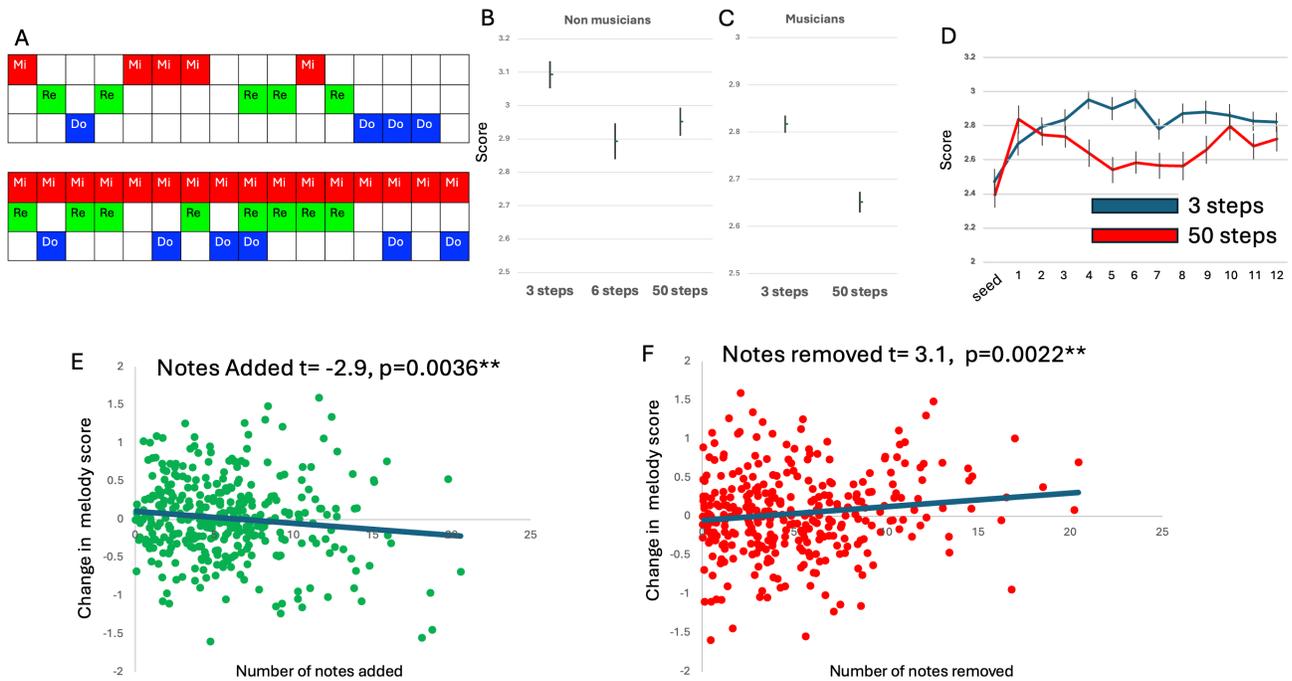

**Figure 2: Non musicians vs. musician cohort**. A, an example of top rated (top) and bottom rated (bottom) tunes. Notes and chords are placed in 16 time slots (see step synthesizer in **Fig. 1A**). Means and s.e.m. of pleasantness scores (validation experiments) for all tunes created over iterations by non-musicians (C) and by musicians (D) for each step size group. E, number of added notes vs. change in rating scores across all participants where steps size = 50. F, number of notes removed by participant vs. change in rating scores, steps size = 50.

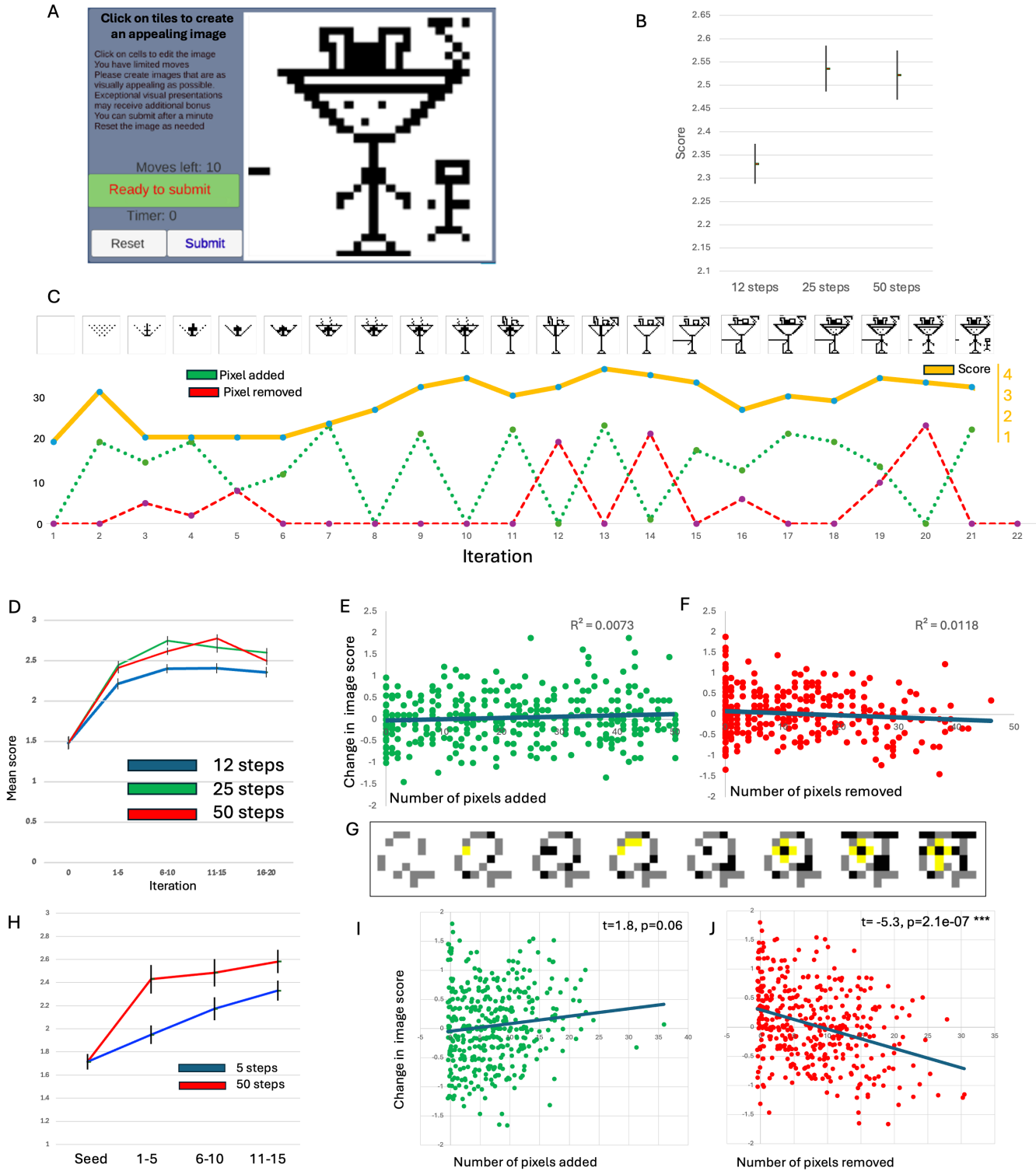

*Figure 3:* Collective creation of images. A, User interface for drawing pixels. B, Image scores (by an independent cohort) over 20 iterations in three groups (XX participants) limiting change to 12, 25 or 50 pixels. C, an example of an evolving chain (25 steps limit). D, time course of changing scores over iterations for each group. E, number of added pixels vs. change in rating scores across all participants where we minimized constraints (50 pixel). F. number of notes removed by participant vs. change in rating scores for the same group. G, Small canvas image, with three colors (w/b/y) and unchangeable pixels (gray). H-J results with the small canvas. Note that here due to the canvas size we adjusted the step limit to 5 vs. 50.

**Evolution of images under different constraints**
Finally, we test if step-size limits affect outcomes of collectively created images. **Figure 3A** presents a user interface canvas (30x30 pixels) we designed for creating simple monochrome images. Experimental design was similar to the previous study: Images evolved in feedforward chains without selection, over 25 iterations, with each participant modifying a single image, which then propagates up the chain. We tested three groups, limiting change to either 12, 25 or 50 pixels (24 chains & 453 participants recruited via Prolific). As shown in **Figure 3B**, we obtained the opposite effect to that of melody creation: images created with a 12 pixel limit were far less appealing than those created by either 25 or 50 step limit (t=5.3 p=1.33e-07 & t=4.95 p=8.31e-07). **Figure 3C** presents an example of an evolving chain as pixels are added and removed. Over iterations, the disadvantage of small step size shows simple dynamics (**Fig. 3D**) with persistent advantage of larger step size. Investigating how the number of added and removed pixels affect pleasantness score revealed no apparent trends (**Fig. 3 E-F,** NS). To further explore for generality, we recruited an additional cohort of participants (n=167, 18 chain), and presented them with a different image creation challenge: The canvas was much smaller (8x8 pixels), the pixels were tri-chromatic, and a random set of 16 pixels in each chain were grayed out and unchangeable, to make the creation of iconic images more difficult (**Fig. 3G**). Due to the small space, we tested a step size of 5 pixels against 50. To sample the space of images evenly, we seeded each chain with random images of 4, 10 or 15 pixels. Each chain was iterated 25 times. Here too, the unlimited step group outperformed the step limit group (**Fig. 3H,** t=-5.167 p=4.45e-07). At the micro level, we found that the number of added and removed pixels affected the score, but as opposed to results with melody, here the score marginally increased with the number of pixels added (t=1.8 p=0.06), and significantly decreased with the number of pixels removed in each step (**Fig 3I-J**, t=-5.3 p= 2.1E-7).

## Discussion

We found that, with melodies, limiting the number of changes in each step resulted in the evolution of more appealing artifacts compared to evolutionary chains where we allowed unlimited freedom to add or remove elements. This effect was replicated several times: with general public recruit and with musicians, with Create & Rate paradigm, and with simple feed-forward (create only) chains. But we failed to replicate this effect with images. With images, cohorts that were given unlimited freedom to add or remove pixels created artifacts that were more appealing compared to cohorts where we limited the number of changes in each step. Here too, this outcome was replicated in different settings.

There are several alternative explanations to these conflicting findings across domains. First, at the micro level, participants who made smaller changes to melodies were more successful in improving the score, an effect that we did not observe with images. Why is that? Creating great melodies and great images are both very difficult. But whereas creating a mediocre melody is inherently hard, most people can create mediocre images. In contrast, any music writing demands coherent association between melody and rhythm, (i.e., rhythmic pattern, beat and meter), and between these parameters and the harmonic structure of the piece. To create a coherent melody one needs an excellent internal representation of pitch and rhythm, and an excellent working memory for sounds. These capacities are acquired over decades of training in professional musicians.

But we suspect that even for melodies, the specific step size effect we observed is not likely to be universal. Our step synthesizer has only 3 notes. It can generate many (~281300 trillion) unique tunes, but musically, we designed it to be limited, with no sub-dominant and only 16 time slots. Therefore, there is a limited number of good outcomes, which again, makes tunes easy to spoil. Beyond that, associative thinking in an evolutionary chain might be more continuous and consistent over iterations in some domains and less in others. For example, working memory, auditory imagination and visual imagination have different constraints in different scenarios (Hubbard, 2013; Stetson, 1896).

The interesting question, therefore, is under what conditions should we expect to see a positive outcome of limiting step size? One option is that evolutionary outcomes might be harder to foresee in some domains and easier in others. We will call this 'visibility' hypothesis. Alternatively, the search 'terrain' itself could be more complicated in some domains, and easier to navigate in others ((Goldenberg et al., 1999; Hart et al., 2017, 2018)). We discuss these hypotheses in the context search algorithms, where these issues were studied intensively. Finally, we will discuss possible parallels in the cultural evolution of art, and in the natural cultural evolution of birdsong, where transmission chains have been studied over decades.

### Parallels with search algorithms

We may think of creativity as a systematic search within a large parameter space (Thompson et al., 2022)(Goldenberg et al., 1999; Hart et al., 2017, 2018); (Thompson et al., 2022). Limiting the number of changes in collective creativity has an interesting parallel with limiting the number of steps in a search algorithm, an optimization problem that has been studied extensively (REFS). Before we can figure out an optimal search strategy, we need to consider the shape of the utility function in a search space. In cases where the utility space is of smooth and simple shape, progressing in big steps can be efficient. In a complex space, however, big steps run the risk of overshooting and reducing utility. It is well established that searching a complex terrain is more efficient in smaller steps (REFS). We are currently investigating if the shape of utility (pleasantness) space is different between images and melodies. Alternatively, our results may be driven not by complexity but by the visibility of the terrain. Specifically, it may be more difficult to predict how a melody will sound without actually making the change, especially for those without musical training or notation skills, compared to visual modifications in images.

### Parallels with birdsong cultures

Male songbirds sing to attract mates (Catchpole & Slater, 2003). Birdsong is composed of syllables that are either improvised or copied from other birds (e.g., father). In many species, song imitation is very accurate. That is, the improvisation to imitation ratio is low, and the majority of syllables produced are copied from other birds (Marler, 1997; Podos & Warren, 2007). However, the imitation vs. improvisation ratio varies strongly across birdsong species. For example the song of the California Thrasher is highly improvised (Sasahara et al., 2012) and we do not understand why. One idea is that, only in certain conditions, a low imitation to improvisation ratio is an optimal evolutionary 'search strategy' for finding the most attractive songs. Testing this idea cannot be conducted directly on birds since this process occurred during the evolutionary time scales. But our collective creativity experiments offer interesting parallels: we can think of people playing the role of creators as male birds, and of those playing the role of evaluator as female birds, in an evolutionary chain. The step limit of three notes, that we found to be most successful, is equivalent to about 20% change given 16 time slots. This is a similar ratio to that observed in the most commonly studied songbirds, including house finches (Ju et al., 2019) and zebra finches (Tchernichovski et al., 2021). The complexity of the songs of these species is also comparable to that of the tunes that can be created with our step synthesizer. It should be feasible to test predictions, by developing synthesizers that match the song complexity of a songbird with a knowed imitation to improvisation ratio, and then measure in human subjects if the optimal step size is similar to that observed in these birds. Using human subjects as a model species for studying cultural evolution in other species (which was never attempted to our knowledge) could perhaps explain some of the natural variability we observe among songbirds.